\documentclass[twoside,twocolumn,9pt]{article}
\usepackage{extsizes}
\usepackage[super,sort&compress,comma]{natbib}
\usepackage[version=3]{mhchem}
\usepackage[left=1.5cm, right=1.5cm, top=1.785cm, bottom=2.0cm]{geometry}
\usepackage{balance}
\usepackage{mathptmx}
\usepackage{sectsty}
\usepackage{graphicx}
\usepackage{lastpage} 
\usepackage[format=plain,justification=justified, singlelinecheck=false,font={stretch=1.125,small,sf},labelfont=bf,labelsep=space]{caption} 
\usepackage{float}
\usepackage{fancyhdr}
\usepackage{fnpos}
\usepackage[english]{babel}

\usepackage{array}
\usepackage{droidsans}
\usepackage{charter}
\usepackage[T1]{fontenc}
\usepackage[usenames,dvipsnames]{xcolor}
\usepackage{setspace}
\usepackage[compact]{titlesec}
\usepackage{hyperref}
\usepackage{physics}
\usepackage{nicefrac}
\usepackage{subcaption}
\usepackage{chemformula}
\usepackage{xspace}

\usepackage{amsmath}
\usepackage{amssymb}
\usepackage{amscd}
\usepackage[np,autolanguage]{numprint}
\usepackage{bm}

\usepackage{xcolor}
\usepackage{hyperref}
\usepackage{cleveref}    % For \cref
\usepackage{braket}      % For \Braket
\usepackage{upgreek}     % For \uppi
\usepackage{chemformula} % For \ch{} 
\usepackage{booktabs}    % Better \toprule \midrule \bottomrule
\definecolor{cream}{RGB}{222,217,201}

\begin{document}

\pagestyle{fancy}
\thispagestyle{plain}
\fancypagestyle{plain}{
%%%HEADER%%%
\renewcommand{\headrulewidth}{0pt}
}
%%%END OF HEADER%%%

%%%PAGE SETUP - Please do not change any commands within this section%%%
\makeFNbottom
\makeatletter
\renewcommand\LARGE{\@setfontsize\LARGE{15pt}{17}}
\renewcommand\Large{\@setfontsize\Large{12pt}{14}}
\renewcommand\large{\@setfontsize\large{10pt}{12}}
\renewcommand\footnotesize{\@setfontsize\footnotesize{7pt}{10}}
\makeatother

\renewcommand{\thefootnote}{\fnsymbol{footnote}}
\renewcommand\footnoterule{\vspace*{1pt}%
\color{cream}\hrule width 3.5in height 0.4pt \color{black}\vspace*{5pt}}
\setcounter{secnumdepth}{5}

\makeatletter
\renewcommand\@biblabel[1]{#1}
\renewcommand\@makefntext[1]%
{\noindent\makebox[0pt][r]{\@thefnmark\,}#1}
\makeatother
\renewcommand{\figurename}{\small{Fig.}~}
\sectionfont{\sffamily\Large}
\subsectionfont{\normalsize}
\subsubsectionfont{\bf}
\setstretch{1.125} %In particular, please do not alter this line.
\setlength{\skip\footins}{0.8cm}
\setlength{\footnotesep}{0.25cm}
\setlength{\jot}{10pt}
\titlespacing*{\section}{0pt}{4pt}{4pt}
\titlespacing*{\subsection}{0pt}{15pt}{1pt}
%%%END OF PAGE SETUP%%%

%%%FOOTER%%%
\fancyfoot{}
\fancyfoot[LO,RE]{\vspace{-7.1pt}\includegraphics[height=9pt]{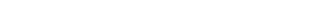}}
\fancyfoot[CO]{\vspace{-7.1pt}\hspace{11.9cm}\includegraphics{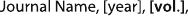}}
\fancyfoot[CE]{\vspace{-7.2pt}\hspace{-13.2cm}\includegraphics{head_foot/RF}}
\fancyfoot[RO]{\footnotesize{\sffamily{1--\pageref{LastPage} ~\textbar  \hspace{2pt}\thepage}}
}
\fancyfoot[LE]{\footnotesize{\sffamily{\thepage~\textbar\hspace{4.65cm} 1--\pageref{LastPage}}
}}
\fancyhead{}
\renewcommand{\headrulewidth}{0pt}
\renewcommand{\footrulewidth}{0pt}
\setlength{\arrayrulewidth}{1pt}
\setlength{\columnsep}{6.5mm}
\setlength\bibsep{1pt}
%%%END OF FOOTER%%%

%%%FIGURE SETUP - please do not change any commands within this section%%%
\makeatletter
\newlength{\figrulesep}
\setlength{\figrulesep}{0.5\textfloatsep}

\newcommand{\topfigrule}{\vspace*{-1pt}%
\noindent{\color{cream}\rule[-\figrulesep]{\columnwidth}{1.5pt}} }

\newcommand{\botfigrule}{\vspace*{-2pt}%
\noindent{\color{cream}\rule[\figrulesep]{\columnwidth}{1.5pt}} }

\newcommand{\dblfigrule}{\vspace*{-1pt}%
\noindent{\color{cream}\rule[-\figrulesep]{\textwidth}{1.5pt}} }

\makeatother
%%%END OF FIGURE SETUP%%%

%%%TITLE, AUTHORS AND ABSTRACT%%%
\twocolumn[
  \begin{@twocolumnfalse}
{\includegraphics[height=30pt]{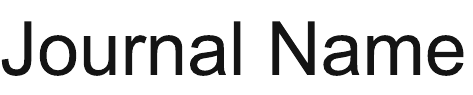}\hfill\raisebox{0pt}[0pt][0pt]{\includegraphics
[height=55pt]{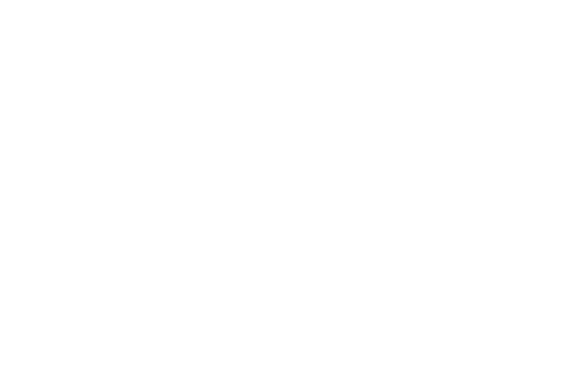}}\\[1ex]
\includegraphics[width=18.5cm]{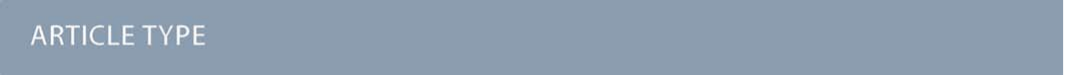}}\par
\vspace{1em}
\sffamily
\begin{tabular}{m{4.5cm} p{13.5cm} }

\includegraphics{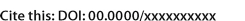} & \noindent\LARGE{\textbf{The exchange-correlation dipole moment dispersion method$^\dag$}} \\
\vspace{0.3cm} & \vspace{0.3cm} \\

%\title{The exchange-correlation dipole moment dispersion method}

 & \noindent\large{Kyle R. Bryenton\textit{$^a$} and Erin R.\ Johnson\textit{$^{a,b,c\ast}$}} \\

\includegraphics{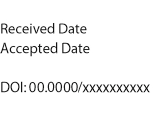} & \noindent\normalsize{
Density-functional theory (DFT) has become the workhorse of modern computational chemistry, with dispersion corrections such as the exchange-hole dipole moment (XDM) model playing a key role in high-accuracy modelling of large-scale systems. Here, we introduce a new physics-guided XDM variant, termed the exchange-correlation dipole moment (XCDM) model, which supplements XDM with same- and opposite-spin dynamical correlation terms, substantially improving accuracy for molecular $C_6$ dispersion coefficients. Both XDM and XCDM are implemented for use with the Becke-Johnson damping function based on atomic radii, as well as a one-parameter damping function based on atomic numbers, recently proposed by Becke. All four variants are benchmarked on the comprehensive GMTKN55 database using minimally empirical generalised-gradient-approximation, global hybrid, and range-separated hybrid functionals. This marks the first time that the XDM (and many-body dispersion, MBD) corrections have been tested for the GMTKN55 set. 
Five solid-state benchmarks spanning molecular crystals and layered materials are also considered. The B86bPBE0 hybrid functional, paired with any of the XDM variants, shows excellent performance for molecular systems. Finally, we identify a flaw in the weighted mean absolute deviation (\mbox{WTMAD-2}) scheme commonly used for the GMTKN55 set, which underweights some of its component benchmarks by orders of magnitude. We propose a new \mbox{WTMAD-4} scheme based on typical errors observed for well-behaved functionals, ensuring fair treatment across all benchmarks.
%\end{abstract}
} \\%The abstract goes here instead of the text "The abstract should be..."

\end{tabular}

 \end{@twocolumnfalse} \vspace{0.6cm}

  ]
%%%END OF TITLE, AUTHORS AND ABSTRACT%%%

%%%FONT SETUP - please do not change any commands within this section
\renewcommand*\rmdefault{bch}\normalfont\upshape
\rmfamily
\section*{}
\vspace{-1cm}

%%%FOOTNOTES%%%

\footnotetext{\textit{$^{a}$~
Department of Physics and Atmospheric Science, Dalhousie University, 6310 Coburg Road, Halifax, Nova Scotia, Canada, B3H 4R2}}

\footnotetext{\textit{$^{b}$~Department of Chemistry, Dalhousie University, 6243 Alumni Crescent, Halifax, Nova Scotia, B3H 4R2, Canada. E-mail: erin.johnson@dal.ca}}

\footnotetext{\textit{$^{c}$~Yusuf Hamied Department of Chemistry, University of Cambridge, Lensfield Road, Cambridge, CB2 1EW, United Kingdom.}}

\footnotetext{\dag~Electronic Supplementary Information (ESI) available, see DOI: 10.1039/cXCP00000x/}

%%%END OF FOOTNOTES%%%

%%%MAIN TEXT%%%%

%\maketitle

\section{Introduction}

Despite being the weakest of the van der Waals forces, London dispersion interactions are collectively extremely important in determining the structural and energetic properties of many chemical systems. Because dispersion physics is not included in most density-functional approximations (DFAs) for modelling electronic structure, they are commonly augmented by a dispersion correction (DC). Numerous such dispersion methods exist in the literature and may be divided into two classes: (i) explicitly non-local corrections that are included within the self-consistent field (SCF) procedure, and (ii) additive, post-SCF corrections. The first type includes the family of van der Waals functionals (vdW-DF),\cite{dion2004van, roman2009efficient, lee2010higher} as well as (r)VV10.\cite{vydrov2010nonlocal,sabatini2013nonlocal} However, due to their non-local nature, these methods are significantly more expensive than additive corrections. Popular post-SCF methods include the Grimme-D series (D1,\cite{grimme2004accurate} D2,\cite{grimme2006semiempirical} D3,\cite{grimme2010consistent} D3BJ,\cite{grimme2011effect} D4\cite{caldeweyher2019generally}); the many-body dispersion family (TS,\cite{tkatchenko2009accurate} MBD@rsSCS,\cite{tkatchenko2012accurate, ambrosetti2014long} MBD-NL\cite{hermann2020density} uMBD,\cite{kim2020umbd} MBD-FI\cite{gould2016fractionally}); and the exchange-hole dipole moment (XDM) model.\cite{johnson2017exchange}

XDM was originally formulated between 2005 and 2007\cite{becke2005density,johnson2006post,becke2007exchange} and has since proven to be one of the most broadly accurate DFA dispersion treatments due to its limited empiricism and inclusion of important physical considerations.\cite{bryenton2023many,bryenton2023oscallot} XDM has demonstrated accuracy, efficiency, and stability in modelling dispersion binding across a highly diverse range of chemical systems, including intermolecular complexes,\cite{kannemann2010van,otero2013non,price2023xdm,nickerson2023comparison} bulk metals,\cite{adeleke2023effects} salts,\cite{otero2020application,christian2021interplay} layered materials,\cite{otero2020asymptotic,rumson2023low} surfaces,\cite{christian2016surface,christian2017adsorption2} and molecular crystals.\cite{otero2012benchmark,otero2014predicting,otero2019dispersion} The recent implementation of XDM in the FHI-aims\cite{blum2009ab} software, and pairing with hybrid functionals, allows computation of molecular crystal lattice energies with the highest accuracy of any dispersion-corrected DFT reported to date.\cite{price2023xdm} It has also shown great success in the area of molecular crystal structure prediction (CSP).\cite{price2023accurate,mayo2024assessment} 

\begin{figure}
\caption{
Sketch (modelled after Ref.~\citenum{buijse2002approximate}) of the exact exchange hole and non-dynamical correlation hole in a diatomic molecule, the sum of which is approximated by the DFA exchange hole. Also shown is the dynamical correlation hole and its sum with the DFA exchange hole to give the DFA exchange-correlation hole. Note that the exchange holes have a normalisation of -1, while the correlation holes have a normalisation of 0.
}
\label{fig:HoleSketch}
\includegraphics[width=\columnwidth]{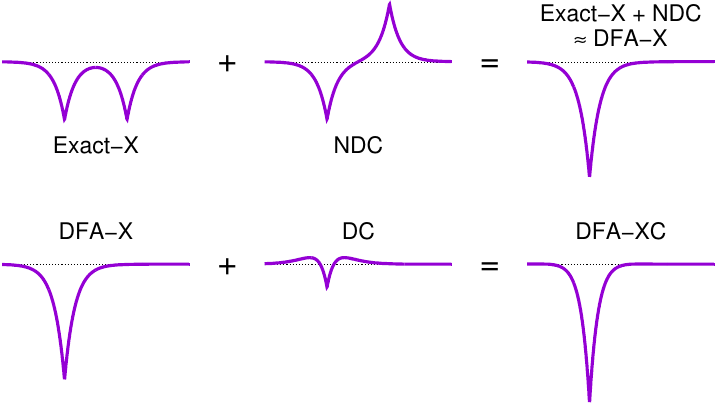}
\end{figure}

The XDM model uses second-order perturbation theory to obtain the dispersion energy in terms of atomic multipole-moment integrals and polarizabilities.\cite{salem1960calculation,dalgarno1966} One key approximation is that the multipole moments are taken to be those of a reference electron and its associated DFA exchange hole,\cite{becke2007exchange} which serves as a convenient and simple proxy for the full exchange-correlation (XC) hole. Indeed, shortly after the formulation of XDM, three papers sought to provide a more rigorous link between the XC hole and the London dispersion energy.\cite{angyan2007exchange, ayers2009perspective, hesselmann2009derivation} The rationale for the good performance of a dispersion model based on the DFA exchange-hole multipole moments, without further correlation terms, is that the chosen Becke--Roussel model\cite{becke1989exchange} is inherently local and confines the hole to a region of roughly atomic size. This implicitly accounts for the effects of non-dynamical correlation (NDC),\cite{becke2003real} which serves to localise the highly non-local exact exchange hole and produce a localised XC hole,\cite{buijse2002approximate} as illustrated in Figure~\ref{fig:HoleSketch}. Thus, NDC is expected to account for the vast majority of the total correlation contribution to the XC hole, while the more local, dynamical correlation (DC) is thought to have only a small contribution\cite{becke2007exchange} and has been neglected in XDM to date. However, given the high accuracy achieved by existing DFA dispersion models, inclusion of the DC contribution is becoming increasingly relevant.

Additionally, Becke recently showed\cite{becke2024remarkably} that XDM fails to accurately predict the binding energies of two alkali-metal clusters (Li$_8$ and Na$_8$) in the ALK8 subset of the GMTKN55 thermochemistry benchmark.\cite{goerigk2017look} The error was traced to the Becke--Johnson (BJ) damping function\cite{johnson2006post} used in XDM (as well as in the D3(BJ) and D4 dispersion models) to damp the dispersion energy to a small negative value at short interatomic separations. An alternative damping function based on atomic numbers, $Z$, was proposed and found to provide good accuracy for these metal clusters, and the GMTKN55 benchmark as a whole.\cite{becke2024remarkably} Notably, the $Z$-dependent damping function is simpler, relying on only one empirical parameter for use with a given DFA, as opposed to the two parameters used in BJ-damping. However, the performance of Z-damping has not yet been assessed on solid-state systems or on molecular systems beyond those comprising the GMTKN55 data set.

In this work, dynamical correlation is added to XDM dispersion through the use of a real-space correlation-hole model,\cite{becke1988correlation,becke1994thermochemical} yielding the exchange-correlation dipole moment (XCDM) dispersion method. The effects of dynamical correlation on the resulting atomic and molecular dispersion coefficients are quantified for the first time. The performance of XDM and XCDM, paired with both BJ- and Z-damping, is assessed for selected isolated-molecule and periodic-solid benchmarks. Overall, XCDM is found to outperform XDM for computation of molecular dispersion coefficients, as well as for all molecular benchmarks, although it significantly overbinds layered materials. XDM with Z-damping appears to be a Pauling point,\cite{lowdin1985twenty} providing consistently reliable results for all benchmarks considered with a minimum of empiricism.

\section{Theory}

\subsection{The XDM Model}

The XDM dispersion energy is written as a sum over all pairs of atoms, $i$ and $j$:
\begin{equation}
E_\text{disp}^\text{XDM} = - \sum_{i<j} \left( \frac{C_{6,ij} f_6}{R^6_{ij}} + \frac{C_{8,ij} f_8}{R^8_{ij}} + \frac{C_{10,ij} f_{10}}{R^{10}_{ij}} \right)\,.
\end{equation}
Here, $C_n$ dispersion coefficients are computed for each atom pair from the self-consistent electron density of the system, as well as the density gradient, Laplacian, kinetic-energy density, and Hirshfeld atomic partitioning weights. The $f_n$ damping functions depend on the interatomic distance, $R_{ij}$, and will be discussed in detail in Section~\ref{ss:damping}.

For a reference electron at coordinate $\bm{r}$, the Pauli exclusion principle necessitates a depletion of probability (and therefore spin density, $\rho_{\sigma}$) of finding a same-spin electron at a nearby reference position $\bm{s}$. This depletion of probability is known as the exchange hole, $h_{\text{X}\sigma}(\bm{r},s)$, for spin $\sigma$. Because the BR hole only requires the angular average of the hole function about the reference point $\bm{r}$, it is thus treated as spherically symmetric with respect to $s$.\cite{becke1988correlation} The reference electron and its exchange hole create a dipole moment given by
\begin{equation} \label{eq:dx}
d_{\text{X}\sigma} (\bm{r}) = \left( \int h_{\text{X}\sigma} \big(\bm{r},s\big) s \, d\bm{s} \right) - \bm{r} \,.
\end{equation}

For a single atom, let the reference electron and exchange-hole centre be located at distances $r$ and $r-d_{\text{X}\sigma}$, respectively, from the nucleus. Then, the strength of that dipole moment would be $r - (r - d_{\text{X}\sigma})=d_{\text{X}\sigma}$. To calculate a higher multipole (or $\ell$-pole) contribution,  the magnitude of those moments at $\bm{r}$ for each spin would be  $r^{\ell} - (r - d_{\text{X}\sigma})^{\ell}$. Thus, the multipole moments for each atom, $i$, in a chemical system are given by
\begin{equation}\label{eq:M}
\Braket{M_{\ell}^{2}}_i = \sum_{\sigma} \int w_{i}(\bm{r}) \, \rho_{\sigma}(\bm{r}) \left[r_{i}^{\ell} - \big(r_{i} - d_{\text{X}\sigma}(\bm{r}) \big)^{\ell} \right]^2 d\bm{r} \,,
\end{equation}
where $r_{i}(\bm{r}) = \left| R_{i} - \bm{r}\right|$ and $R_{i}$ is the position of atom $i$. In practice, $r_i-d_{\text{X}\sigma}(\bm{r})$ is enforced to be $\ge 0$ as the exchange-hole dipole moment should not exceed the distance from the reference electron to the nearest nucleus.\cite{becke2007unified}
The partitioning into atomic contributions is achieved by inclusion of the Hirshfeld weights, $w_i(\bm{r})$, in the integrand.\cite{hirshfeld1977bonded, heidar2017information}

To calculate the $C_n$ dispersion coefficients, the multipole moment integrals are combined with atom-in-molecule polarizabilities, 
\begin{equation} \label{eq:alpha}
\alpha_{i} = \alpha_{i}^{\text{free}} \frac{ v_{i} }{ v_{i}^{\text{free}} } \,,
\end{equation}
where $\alpha^{\text{free}}$ is the free atomic polarizability taken from readily available sources,\cite{rumble2021crc} and $v_{i} = \Braket{r^{3}}_{i}$ is the atom-in-molecule volume computed using the Hirshfeld partitioning of the electron density. The dispersion coefficients are then given by
\begin{align}
C_{6,ij} &= \frac{\alpha_{i} \, \alpha_{j} \Braket{M_{1}^{2}}_{i} \Braket{M_{1}^{2}}_{j} }{\alpha_{i}\Braket{M_{1}^{2}}_{j} + \alpha_{j} \Braket{M_{1}^{2}}_{i}} \,, \\
C_{8,ij} &= \frac{3}{2} \frac{\alpha_{i} \, \alpha_{j} \left( \Braket{M_{1}^{2}}_{i} \Braket{M_{2}^{2}}_{j} + \Braket{M_{2}^{2}}_{i} \Braket{M_{1}^{2}}_{j} \right)}{\alpha_{i}\Braket{M_{1}^{2}}_{j} + \alpha_{j} \Braket{M_{1}^{2}}_{i}} \,, \\
C_{10,ij} &= 2 \frac{\alpha_{i} \, \alpha_{j} \left( \Braket{M_{1}^{2}}_{i} \Braket{M_{3}^{2}}_{j} + \Braket{M_{3}^{2}}_{i} \Braket{M_{1}^{2}}_{j} \right)}{\alpha_{i}\Braket{M_{1}^{2}}_{j} + \alpha_{j} \Braket{M_{1}^{2}}_{i}} \notag\\
& \quad + \frac{21}{5} \frac{\alpha_{i} \, \alpha_{j} \Braket{M_{2}^{2}}_{i} \Braket{M_{2}^{2}}_{j}}{\alpha_{i}\Braket{M_{1}^{2}}_{j} + \alpha_{j} \Braket{M_{1}^{2}}_{i}} \,. 
\end{align} 
Clearly, $C_{6,ij}$ accounts for dipole-dipole, $C_{8,ij}$ accounts for dipole-quadrupole, and $C_{10,ij}$ accounts for both the quadrupole-quadrupole and dipole-octupole contributions to the dispersion interaction.

In XDM's original formulation,\cite{becke2005exchange} the dipole moments in Eq.~\ref{eq:dx} used the exact exchange hole, written in terms of the occupied Kohn--Sham orbitals. This was later changed\cite{becke2005density} to use the Becke--Roussel (BR) hole,\cite{becke1989exchange} which has the benefits of being much less computationally demanding and more localized to an atomic-sized region than the exact exchange hole. 
As previously stated, this localisation mimics the effect of non-dynamical correlation that arises from chemical bonding, allowing improved accuracy when applying XDM to molecular systems.\cite{becke2007unified}

The BR hole is formally a meta-generalised-gradient-approximation (meta-GGA) functional that models 
the exchange hole as an exponential function of the form $A\textrm{e}^{-ar}$ centred a distance $b$ from its reference electron. To determine the values of the three parameters $(A_\sigma,a_\sigma,b_\sigma)$ for a given position and spin ($\sigma$) of the reference electron, three requisite constraints are imposed. Firstly, the model hole must be normalized to -1 electron, thus 
\begin{equation}
A_\sigma = -\frac{a_{\sigma}^3}{8 \uppi} \,.
\end{equation}
Second, the model hole must deplete to the spin density at the reference point, meaning that
\begin{equation}\label{eq:rhobr}
\rho_{\sigma} = \frac{a_{\sigma}^3}{8 \uppi} \textrm{e}^{-a_{\sigma}b_{\sigma}} \,.
\end{equation}
Third, the model hole must have the same curvature as the exact exchange hole at the reference point, given by
\begin{equation}
Q_{\sigma} = \frac{1}{6} \left[ \nabla^{2} \rho_{\sigma} - 2 \tau_{\sigma} + \frac{1}{2} \frac{ \left(\nabla \rho_{\sigma} \right)^2}{\rho_{\sigma}} \right] \,,
\end{equation}
where
\begin{equation}\label{eq:tau}
\tau_{\sigma} = \sum_{i} \left| \nabla \psi_{i\sigma}\right|^2
\end{equation}
is the kinetic energy density, written in terms of occupied Kohn--Sham orbitals, $\psi_{i\sigma}$, and following Becke's notation where the usual $\nicefrac{1}{2}$ factor is omitted. Substituting our BR spin density into this formula yields a hole curvature of
\begin{equation}
Q_{\sigma} = \frac{\rho_{\sigma}}{6 b_{\sigma}} \left[ a_{\sigma}^{2} b_{\sigma} - 2 a_{\sigma} \right] \,.
\end{equation}
We can then solve for the exponent, $a_\sigma$, and hole displacement, $b_\sigma$, by letting $x=a_\sigma b_\sigma $ and solving the transcendental equation
\begin{equation}
\frac{x \, \textrm{e}^{-2x/3}}{x-2} = \frac{2}{3} \uppi^{2/3} \frac{\rho_{\sigma}^{5/3}}{Q_{\sigma}} \,.
\end{equation}
This equation is typically solved iteratively using the Newton--Raphson method. Solving for $x$ lets us analytically determine $b_{\sigma}$, and thus $a_{\sigma}$, via a rearrangement of Eq.~\ref{eq:rhobr} as
\begin{equation}
b_{\sigma}^{3} = \frac{x^{3} \textrm{e}^{-x}}{8 \uppi \rho_{\sigma}} \,.
\end{equation}
Lastly, the exchange-hole dipole moment is taken as the distance between a reference electron and the centre of the exchange hole, thus $d_{\text{X}\sigma} = b_{\sigma}$. 

\subsection{Inclusion of Dynamical Correlation: XCDM}\label{ss:xcdm}

As previously stated, since XDM uses the BR hole model it is able to capture the effects of non-dynamical correlation as well as exchange. However, the BR model does not account for dynamical correlation, so it must be added. This is intended; it has long been argued that the role of DFA exchange functionals is to also capture this non-dynamical contribution to correlation, and that dynamical correlation should be captured explicitly through dedicated correlation functionals.\cite{neumann1996exchange, becke1994thermochemical} We consider the same-spin ($\sigma\sigma$) and opposite-spin ($\sigma\sigma^\prime$ for $\sigma \ne \sigma^\prime$) dynamical correlation holes proposed by Becke:\cite{becke1988correlation}
\begin{align}\label{eq:choles}
h_{\text{C}\sigma\sigma} (\bm{r},s) &= \frac{s^2 \left( s - z_{\sigma\sigma} \right) D_{\sigma}(\bm{r})}{6 \left(1 + z_{\sigma\sigma}/2 \right)}  F\left(\gamma_{\sigma\sigma} s \right) \,, \\
h_{\text{C}\sigma\sigma^\prime} (\bm{r},s) &= \frac{ \left( s - z_{\sigma\sigma^\prime} \right) \rho_{\sigma^\prime} (\bm{r})}{1 + z_{\sigma\sigma^\prime}} F\left(\gamma_{\sigma\sigma^\prime} s \right) \,.
\end{align}
Here, $z$ is the correlation length, which is the radial distance from the reference electron at which the correlation hole becomes zero. This length is determined using the inverse of the spin-indexed exchange potentials,
\begin{align}
z_{\sigma\sigma} &= 2 c_{\sigma\sigma}  |U_{\text{X}\sigma}|^{-1} \,, \\
z_{\sigma\sigma^\prime} &= c_{\sigma\sigma^\prime} \left( |U_{\text{X}\sigma}|^{-1} + |U_{\text{X}\sigma^\prime}|^{-1} \right) \,,
\end{align}
where the dimensionless quantities $c_{\sigma\sigma}=0.63$ and $c_{\sigma\sigma^\prime}=0.88$ were obtained from fits to atomic correlation energies.\cite{becke1994thermochemical} For the case of the BR exchange functional, the exchange potential is\cite{becke1989exchange} 
\begin{equation}
|U_{\text{X}\sigma}| = \frac{1}{b_\sigma} \left( 1 - \textrm{e}^{-x} - \frac{1}{2} x \, \textrm{e}^{-x} \right) \,.
\end{equation}
Returning to Eq.~\ref{eq:choles}, $D_{\sigma} = \tau_{\sigma} - \tau_{\sigma}^{w}$ is the difference between the exact kinetic-energy density and the von Weizs\"{a}cker approximation,\cite{weizsacker}
\begin{equation}
\tau_{\sigma}^{w} = \frac{1}{4} \frac{\left( \nabla \rho_{\sigma} \right)^2}{\rho_{\sigma}} \,,
\end{equation}
where the prefactor is $\nicefrac{1}{4}$, as opposed to $\nicefrac{1}{8}$, to be commensurate with our definition of $\tau_\sigma$ in Eq.~\ref{eq:tau}. Finally, $F(\gamma s)$ is a function that describes the hole shape and involves an adjustable parameter to enforce normalisation.

In his work, Becke suggested three forms for this normalisation function:\cite{becke1988correlation}
\begin{align} 
F_{\text{1}}(x) &= \textrm{sech}(x) \,, \label{eq:f1}\\
F_{\text{2}}(x) &= \left( 1 + x \right) \textrm{e}^{-x} \,, \label{eq:f2}\\
F_{\text{3}}(x) &= \textrm{e}^{-x^2} \,. \label{eq:f3}
\end{align}
If the exchange and exchange-correlation holes both normalize to -1 electrons, then the correlation hole must normalize to zero. Thus, the first step is to solve for the value of $\gamma$ in these functions to enforce the zero normalisation constraint of
\begin{equation}
\int h_{\text{C}} (\bm{r},s) s^2 \sin(\theta) \, ds \, d\theta \, d\phi = 0 \,.
\end{equation}
For the same-spin hole, it can be shown that
\begin{align}
\gamma_{\sigma\sigma,1} &= \frac{\Psi^{(5)}(\nicefrac{1}{4}) - \Psi^{(5)}(\nicefrac{3}{4})}{360 \uppi^5 z_{\sigma\sigma}} \,,\\
\gamma_{\sigma\sigma,2} &= \frac{35}{6 \, z_{\sigma\sigma}} \,,\\
\gamma_{\sigma\sigma,3} &= \frac{8}{3 \sqrt{\uppi} \, z_{\sigma\sigma}} \,,
\end{align}
and for the opposite-spin hole
\begin{align}
\gamma_{\sigma\sigma^\prime,1} &= \frac{\Psi^{(3)}(\nicefrac{1}{4}) - \Psi^{(3)}(\nicefrac{3}{4})}{16 \uppi^{3} z_{\sigma\sigma^\prime}}\,,\\
\gamma_{\sigma\sigma^\prime,2} &= \frac{15}{4 \, z_{\sigma\sigma^\prime}}\,,\\
\gamma_{\sigma\sigma^\prime,3} &= \frac{2}{\sqrt{\uppi} \, z_{\sigma\sigma^\prime}}\,,
\end{align}
where $\Psi^{(n)}(z) = \frac{d^{n}}{dz^{n}} \frac{\Gamma'(z)}{\Gamma(z)}$ is the polygamma function, not to be confused with a wavefunction. Then, evaluated analogously to Eq.~\ref{eq:dx}, our same- and opposite-spin correlation-hole dipole moment contributions are
\begin{align}
d_{\text{C}\sigma\sigma}(\bm{r})  &= \left[ \frac{g_{\sigma\sigma} \, z_{\sigma\sigma}^{7}}{2 + z_{\sigma\sigma}} D_{\sigma}(\bm{r}) \right] - \bm{r}\,, \\
d_{\text{C}\sigma\sigma^\prime}(\bm{r})  &= \left[ \frac{g_{\sigma\sigma^\prime} \, z_{\sigma\sigma^\prime}^{5}}{1 + z_{\sigma\sigma^\prime}} \rho_{\sigma^\prime}(\bm{r}) \right] - \bm{r}\,,
\end{align}
where the value of the dimensionless constant $g$ depends on the chosen normalisation function, $F$, and is thus a function of $\gamma$. XCDM was tested using each normalisation function on a molecular $C_6$ benchmark (see Section~\ref{ss:c6}), and the values from the sech-type form of Eq.~\ref{eq:f1} were ultimately chosen, specifically $g_{\sigma\sigma} = 0.01243$ and $g_{\sigma\sigma^\prime} = 0.5360$. 

Finally, the correlation-hole contributions are combined with the exchange-hole dipole contribution to form the $\sigma$-spin exchange-correlation hole, given by 
\begin{equation}
h_{\text{XC}\sigma} (\bm{r},s) = h_{\text{X}\sigma} (\bm{r},s) + h_{\text{C}\sigma\sigma} (\bm{r},s) + h_{\text{C}\sigma\sigma'} (\bm{r},s) \,.
\end{equation}
The exchange-correlation hole dipole moment is then given by
\begin{align}\label{eq:dxc}
d_{\text{XC}\sigma} &= \left( \int h_{\text{XC}\sigma} \,  s \, d\bm{s} \right) - \bm{r} \notag\\
&= \left( \int \left[ h_{\text{X}\sigma} + h_{\text{C}}^{\sigma\sigma}  + h_{\text{C}}^{\sigma\sigma'} \right]  s \, d\bm{s} \right) - \bm{r} \notag\\
&= b_{\sigma}  + \left[ \frac{g_{\sigma\sigma} \, z_{\sigma\sigma}^{7}}{2 + z_{\sigma\sigma}} D_{\sigma} \right] + \left[ \frac{g_{\sigma\sigma'} \, z_{\sigma\sigma'}^{5}}{1 + z_{\sigma\sigma'}} \rho_{\sigma'} \right] -  \bm{r} \,,
\end{align} 
which is substituted into the multipole-moment integrals of Eq.~\ref{eq:M} in order to evaluate the XCDM dispersion coefficients.

\begin{figure}
\includegraphics[width=\columnwidth]{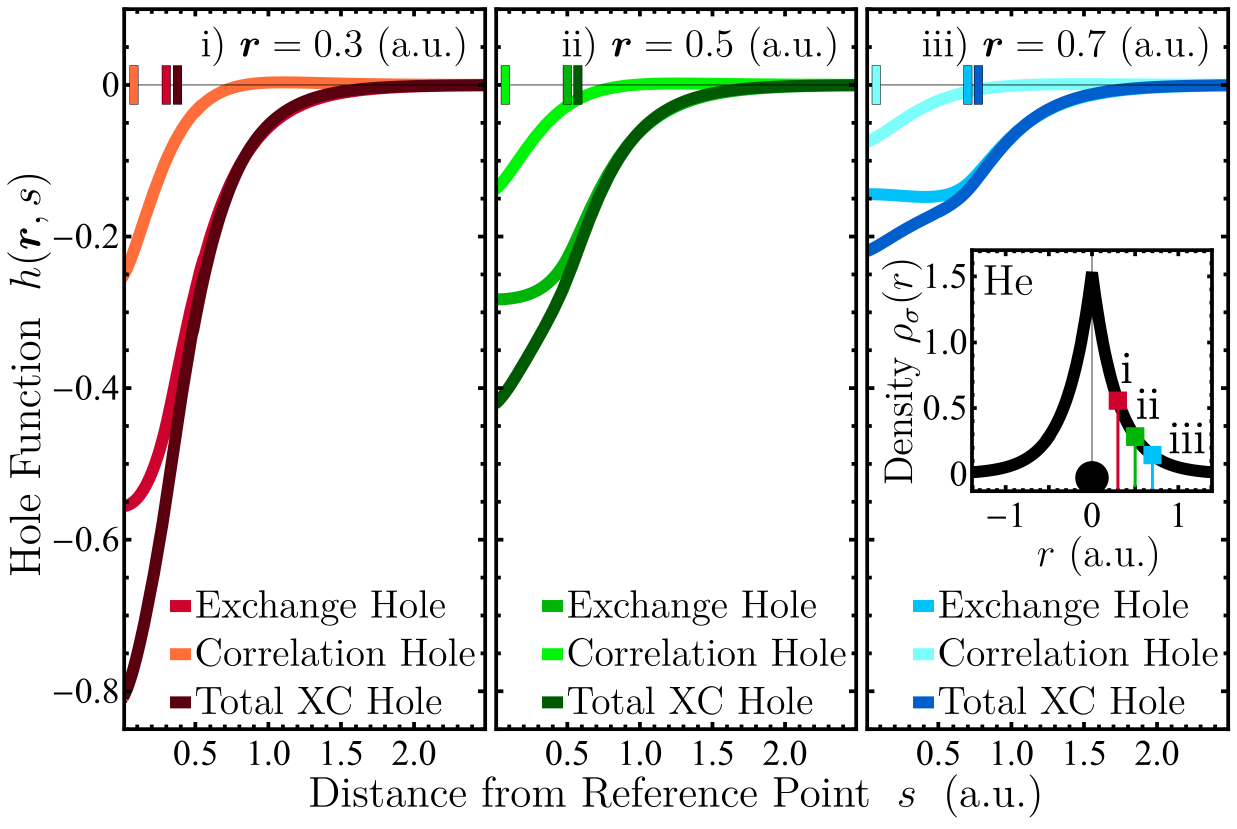}
\caption{The exchange hole, opposite-spin correlation hole, and combined XC hole are plotted as a function of the distance from the reference electron, $s$. The associated dipole moment strengths are indicated with coloured bars along the $s$-axis. Results are shown for the helium atom, with the reference electron at three selected displacements, $r$ (in bohr), from the nucleus, as shown in the inset.}
\label{fig:xchole}
\end{figure}

Figure~\ref{fig:xchole} shows the radial distribution of the exchange hole, correlation hole, and total XC hole about a reference electron for three positions within an isolated helium atom. Naturally, all three holes are deeper when the reference electron is closer to the nucleus, as the overall electron density is higher. Addition of dynamical correlation deepens the overall hole near the reference point, but its effect is rather short range and limited by the correlation length, which is ca.~1 a.u.\@ in this particular example. The figure also shows the magnitudes of the exchange-, correlation-, and XC-hole dipole moments. Due to the localised nature of the dynamical correlation hole about the reference electron, it gives rise to a small dipole moment. Thus, the XC-hole dipole moment is only slightly larger than the exchange-hole dipole moment, justifying the previous neglect of the dynamical correlation contribution in XDM. However, the dynamical correlation contribution is non-negligible, and its inclusion does result in a ca.~10\% decrease in the mean absolute percent error (MAPE) in molecular $C_6$ dispersion coefficients. Further, as shown in Table~\ref{tab:molc6}, the XCDM mean percent error (MPE) all but vanishes, eliminating the systematic underestimation of the molecular $C_6$ dispersion coefficients observed with XDM. Tabulated MAPE and MPE values for the $F_2$ and $F_3$ normalisation functions, as well as the rounded mean of all three normalisation functions, can be found in the ESI.

\begin{table}[t]
\caption{Results for the MolC6 benchmark of homomolecular $C_6$ coefficients, computed using XDM and XCDM in FHI-aims\cite{blum2009ab} using \texttt{tight} basis settings. Mean percent errors (MPE) and mean absolute percent errors (MAPE) are shown for the B86bPBE and PBE functionals, as well as their 25\% and 50\% hybrid variants.}
\label{tab:molc6}
\centering
\begin{tabular}{l|cc|cc}\hline
\multicolumn{1}{c|}{} & \multicolumn{2}{c|}{XDM} & \multicolumn{2}{c}{XCDM} \\
Functional&    MAPE   &  MPE    &    MAPE   &  MPE   \\ \hline
B86bPBE   &    18.4   & -16.4   &    8.5    & ~2.7   \\
B86bPBE0  &    19.6   & -18.1   &    8.6    & ~0.6   \\
B86bPBE50 &    17.0   & -14.5   &    9.0    & -1.2   \\
PBE       &    18.1   & -16.1   &    8.4    & ~3.3   \\
PBE0      &    19.4   & -17.9   &    8.5    & ~1.1   \\
PBE50     &    16.6   & -14.0   &    8.9    & -0.9   \\ \hline
\end{tabular}
\end{table}

\subsection{Damping Functions}\label{ss:damping}

Conventionally, XDM uses the Becke--Johnson (BJ) damping function,\cite{johnson2006post} which is also used in the D3(BJ) and D4 dispersion methods of Grimme and co-workers. The BJ-damping function is
\begin{equation}
f_n^\text{BJ}(R_{ij}) = \frac{R^n_{ij}}{R^n_{ij} + R^n_{\text{vdW},ij}}
\end{equation}
where $R_{\text{vdW},ij}$ is the sum of approximate van der Waals radii of atoms $i$ and $j$. It is determined as
\begin{equation}
R_{\text{vdW},ij} = a_1 R_{c,ij} + a_2
\end{equation}
where $a_1$ and $a_2$ are empirical parameters that are not element-dependent but are fitted for use with a particular combination of density functional and basis set. $R_{c,ij}$ is a ``critical'' interatomic distance at which successive terms in the perturbation theory expansion of the dispersion energy become equal. If the dispersion energy only includes the $C_6$ and $C_8$ terms, then 
\begin{equation}
R_{c,ij} = \sqrt{\frac{C_{8,ij}}{C_{6,ij}}} \,.
\end{equation}
However, if the $C_{10}$ term is also included in the dispersion energy, two other possible definitions for $R_{c,ij}$ arise:
\begin{equation}
R_{c,ij} = \left\{ \begin{array}{l} \sqrt{\frac{C_{10,ij}}{C_{8,ij}}} \\[8pt] \sqrt[4]{\frac{C_{10,ij}}{C_{6,ij}}} \end{array} \right.
\end{equation}
In XDM, the value of $R_{c,ij}$ is taken to be the average of these three results.

Becke recently proposed an alternative damping function for use with XDM that, unlike BJ damping, involves only one empirical fit parameter.\cite{becke2024remarkably} In this work, it will be referred to as Z-damping, due to the dependence on the atomic number. The Z-damping function is 
\begin{equation}\label{eq:zdamp}
f_n^\text{Z} (R_{ij}) = \frac{R^n_{ij}}{R^n_{ij} + z_\text{damp}
\frac{C_{n,ij}}{Z_i+Z_j}}
\end{equation}
where $Z_i$ and $Z_j$ are the atomic numbers of atoms $i$ and $j$, respectively. This definition was chosen because the resulting contribution to the correlation energy in the united-atom limit would be
\begin{equation}
\lim_{R_{ij}\rightarrow0} \left( \frac{C_{n,ij}}{R_{ij}^n + z_\text{damp} \frac{C_{n,ij}}{Z_i+Z_j}} \right) =  \frac{Z_i + Z_j}{z_\text{damp}} 
\end{equation}
and atomic correlation energies are roughly proportional to atomic number. Similar to BJ damping, the single empirical parameter, $z_\text{damp}$, is atom-independent and fitted for use with a particular density functional and basis set. A typical value of $z_\text{damp}$ is around $10^{5}~\text{Ha}^{-1}$.

\begin{figure}
\includegraphics[width=\columnwidth]{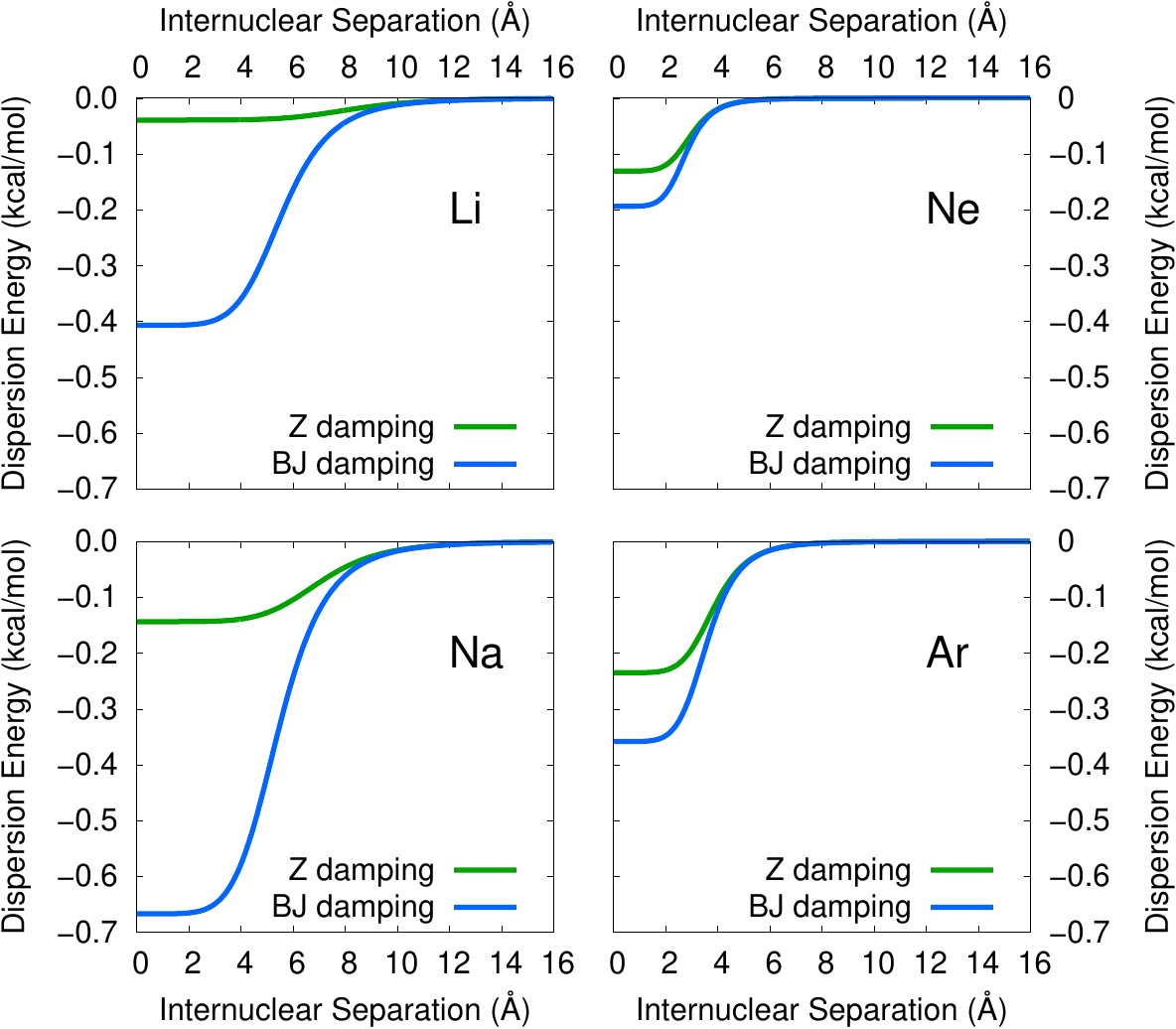}
\caption{Comparison of BJ- and Z-damping functions. The plots use XDM data for the free atoms only, computed with the B86bPBE functional and \texttt{tight} basis settings using FHI-aims.}
\label{fig:damp}
\end{figure}

To illustrate the differences in damping functions, BJ- and Z-damping are compared for homonuclear interactions between Li, Na, Ne, and Ar atoms in Figure~\ref{fig:damp}. The Li and Na calculations are spin-polarized, with a net spin of 1 electron. For simplicity, the curves use only data for the free atoms, which omit changes in dispersion coefficients with internuclear separation that would be observed in the dimer systems due to varying electron densities. The results in Figure~\ref{fig:damp} show that Z-damping consistently reduces the magnitude of the dispersion energy compared to BJ-damping. However, this effect is fairly minor for Ne and Ar, while there is a very large increase in damping strength for Li and Na. This allows correction of the overbinding seen with BJ-damping for the Li$_8$ and Na$_8$ clusters, while preserving high accuracy for main-group elements. With BJ-damping, the magnitudes of the dispersion energies in the united-atom limit follow the trend Na$>$Li$>$Ar$>$Ne, but this changes to Ar$>$Na$\approx$Ne$>$Li for Z-damping. The latter appears more physical because, in the united-atom limit, the dispersion energy would become a correlation energy and should increase with the number of electrons and, hence, atomic number. 

\section{Data Sets}

To evaluate the performance of XDM and XCDM, both with BJ- and Z-damping, a comprehensive list of benchmarks has been selected for testing. These are categorised into three groups: those used to optimize parameters for dispersion coefficients and damping functions, finite-molecule benchmarks, and periodic-solid benchmarks. The benchmark content, geometry sources, and reference data quality are summarised below.

\subsection{$C_6$ Benchmark and Fit Set}\label{ss:c6}

\textbf{MolC6}:
Benchmark of 20 homomolecular $C_6$ dispersion coefficients originally studied by Becke and Johnson.\cite{johnson2005post} Systems include \ch{H2}, \ch{N2}, \ch{O2}, \ch{Cl2}, \ch{CO2}, methane, \ch{CCl4}, \ch{SiH4}, \ch{SiF4}, \ch{SF6}, ethyne, ethene, ethane, propene, propane, butene, butane, pentane, hexane, and benzene. Reference values were obtained from experimental dipole oscillator strength (DOS) data and subsequently refined using the dipole oscillator strength distribution (DOSD) method.\cite{johnson2005post, spackman1991time, kumar1996reliable, kumar1992dipole, kumar2002reliable, kumar2003dipolea, kumar2003dipoleb, wu2002empirical} These $C_6$ reference values were used to assess the accuracy of XCDM relative to XDM and to guide the optimal determination of the $g_{\sigma\sigma}$ and $g_{\sigma\sigma^\prime}$ parameters, as described in Section~\ref{ss:xcdm}.

\textbf{KB49}:
Binding energies of 49 molecular dimers with reference values from basis-set extrapolated CCSD(T) calculations.\cite{kannemann2010van} Dimer geometries are available from the \texttt{refdata} GitHub repository.\cite{otero2015refdata} The BJ-damping parameters, $a_1$ and $a_2$ (in \AA), as well as the Z-damping parameter, $z_\text{damp}$, were fitted separately for XDM and XCDM for each combination of DFA and basis set. Optimal parameters were determined by minimising the root-mean-square percent error (RMSPE) for the KB49 set and may be found in the ESI. 

\subsection{Molecular Benchmarks}

\textbf{GMTKN55}: A collection of 55 individual benchmarks spanning the thermochemistry of small and large molecules, reaction barriers, and both intramolecular and intermolecular non-covalent interactions (NCI). Typically, error metrics are reported for seven categories: ``Basic + Small'', ``Iso + Large'', ``Barriers'', ``Intermolecular NCI'', ``Intramolecular NCI'', ``All NCI'', and GMTKN55 as a whole. The interested reader is directed to Ref.~\citenum{goerigk2017look} for detailed information regarding the individual benchmarks. Geometries for FHI-aims may be obtained from the \texttt{gmtkn55-fhiaims} GitHub repository.\cite{johnson2024gmtkn55} 

Due to the wide range of energy scales of the component benchmarks within the GMTKN55, the overall error for the set is reported as a weighted mean absolute deviation (WTMAD). Several definitions for such a weighted error have been proposed. The first proposed metric, \mbox{WTMAD-1}, is not commonly used. In this scheme, each subset is assigned an arbitrary weight, denoted $w_{i}$, where $w_{i}=10$ when  $\overline{|\Delta E|}_{i} < 7.5 \text{ kcal/mol}$, $w_{i}=0.1$ when  $\overline{|\Delta E|}_{i} > 75 \text{ kcal/mol}$, and $w_{i}=1$ otherwise. The \mbox{WTMAD-1} is then calculated as
\begin{equation} \label{eq:wtmad1}
\text{WTMAD-1} = \frac{1}{N_\text{bench}} \sum_{i=1}^{N_\text{bench}} w_{i} \cdot \text{MAD}_{i} \,.
\end{equation}
The most widely used metric for the GMTKN55 benchmark is the \mbox{WTMAD-2}, introduced in Ref.~\citenum{goerigk2017look} and defined as  
\begin{equation}
\text{WTMAD-2} = \sum_{i=1}^{N_\text{bench}} \frac{N_{i}}{N_{\text{total}}} \cdot \frac{ {\overline{|\Delta E|}_{\text{mean}}}  }{\overline{|\Delta E|}_{i}} \cdot \text{MAD}_{i} \,,
\end{equation}
where the sum runs over all 55 benchmarks. Here, $N_i$ is the number of data points in the $i$th benchmark, $\overline{|\Delta E|}_{i}$ is the average reference energy for that benchmark, $N_{\text{total}} = \sum_{j=1}^{N_\text{bench}} N_{j}$,  $\overline{|\Delta E|}_{\text{mean}}$ is the average of all $\overline{|\Delta E|}_{i}$ values (approximately 56.84 kcal/mol if all 55 subsets are considered), and MAD$_i$ is the mean absolute deviation between the computed and reference data. Additionally, Whittmann \textit{et al.\@} recently proposed the \mbox{WTMAD-3} weighting scheme, which is identical to \mbox{WTMAD-2} except that it attenuates the weights for a subset to be no more than 1\% of the total reactions considered.\cite{wittmann2023dispersion}

As shown in the ESI, all three of these metrics result in a number of benchmarks having a disproportionately large contribution to the overall WTMAD, while others have near-zero contribution. Upon review, we determined that calculating weights based on the MAD relative to the reference energy $\overline{|\Delta E|}_{i}$ was not representative. For example, IL16's average reference energy is 109.04 kcal/mol---171 times larger than its mean MAE across our methods of 0.63 kcal/mol---resulting in severe undercontribution (0.05\%) in \mbox{WTMAD-2}. Subsets such as PA26 (0.21\%) and DIPCS10 (0.04\%) are similarly affected. There also exist cases where the inverse is true and the ratio is small, causing overcontribution of subsets such as PCONF21 (4.5\%), HEAVY28 (4.7\%), and BH76 (9.6\%).

To address these issues, we propose yet another metric, \mbox{WTMAD-4}. This scheme is identical to \mbox{WTMAD-1} in its construction, but the weights are based on the magnitudes of expected errors rather than on the absolute energy scales. As a result, each benchmark contributes meaningfully and appropriately to the overall \mbox{WTMAD-4}, with contributions ranging approximately between 1.0 to 3.5\%. The weights are given by
\begin{equation}
w_i = \left\{ \begin{array}{cl}
50  & \text{ACONF, RG18}\\[4pt]
25  & \text{ADIM6, Amino20x4, BUT14DIOL,}\\
    & \text{HEAVY28, ICONF, MCONF, S66}\\[4pt]
10  & \text{BHROT27, HAL59, IL16, PCONF21,}\\
    & \text{PNICO23, RSE43, S22, SCONF, UPU23}\\[4pt]
5   & \text{AHB21, CARBHB12, CDIE20,}\\
    & \text{CHB6, ISO34, PArel, TAUT15}\\[4pt]
2.5 & \text{AL2X6, BH76, BH76RC, BHPERI,}\\
    & \text{BSR36, FH51, G21EA, HEAVYSB11,}\\
    & \text{IDISP, INV24, ISOL24, NBPRC,}\\
    & \text{PA26, YBDE18}\\[4pt]
1   & \text{ALK8, ALKBDE10, BHDIV10, DARC,}\\
    & \text{DIPCS10, G21IP, G2RC, PX13,}\\
    & \text{RC21, W4‐11, WATER27, WCPT18}\\[4pt]
0.5 & \text{C60ISO, DC13, MB16-43, SIE4x4}\\[4pt]
\end{array} \right.
\end{equation}
Considerations for each subset, such as its total fraction of reactions  within its category, and the quality of its reference data, were taken into account when determining the appropriate weights. For additional information on the reasoning and construction of \mbox{WTMAD-4}, the reader is directed to the ESI.

\subsection{Solid-State Benchmarks}

\textbf{X23}:  %\cite{reilly2013understanding, otero2012benchmark, dolgonos2019revised}}
Lattice energies of 23 molecular crystals,\cite{reilly2013understanding,otero2012benchmark} using updated ``X23b'' reference energies.\cite{dolgonos2019revised}  Geometries are available from the \texttt{refdata} repository.\cite{otero2015refdata} Unlike the previous benchmarks, X23 requires geometry optimisations with each functional and basis combination considered. 

\textbf{HalCrys4}:  %\cite{otero2019dispersion, dean1999lange}}
Lattice energies of four halogen crystals---\ch{Cl2}, \ch{Br2}, \ch{I2}, and \ch{ICl}.\cite{otero2019dispersion} The lattice energies are compared to back-corrected experimental results from Ref.~\citenum{dean1999lange}. As with the X23, geometries are optimised for each reported functional and basis set. Geometries are available from the \texttt{refdata} repository.\cite{otero2015refdata}

\textbf{ICE13}:  
Absolute lattice energies of ice polymorphs \cite{brandenburg2015benchmarking}
(Abs), along with their relative energy differences (Rel) using diffusion Monte Carlo (DMC) reference data.\cite{della2022dmc} ICE13 requires geometry optimisations for all systems except the isolated water molecule, which uses a fixed geometry. Geometries are available from the \texttt{refdata} repository.\cite{otero2015refdata}

\textbf{LM26}: %\cite{bjorkman2014testing, tawfik2018evaluation}}
Exfoliation energies of 26 layered materials, predominantly transition-metal dichalcogenides, but also including graphite and hexagonal boron nitride.\cite{bjorkman2014testing} We also provide statistics for the LM11 subset studied by Tawfik.\cite{tawfik2018evaluation} Reference data was obtained using the random-phase approximation (RPA), and geometries may be obtained from the Inorganic Crystal Structure Database.\cite{zagorac2019recent} Since small deviations in the in-plane lattice constant, $a$, do not significantly affect the equilibrium interlayer separation and binding energy, the in-plane lattice constants are fixed to their experimental values. The binding energies and $c$ lattice constant are determined by unit-cell relaxation\cite{tawfik2018evaluation} or interpolation.\cite{bjorkman2014testing, bryenton2023many} 

\section{Computational Methods} \label{ss:compmethods}

Unless otherwise stated, all calculations were performed using version 250425 of FHI-aims (commit b38a7049).\cite{blum2009ab, ren2012resolution, levchenko2015hybrid, kokott2024efficient, yu2018elsi, havu2009efficient, ihrig2015accurate, price2023xdm}
As noted above, the BJ- and Z-damping coefficients were determined for each functional and basis combination by least-squares fitting to minimise the RMSPE for the KB49 benchmark set of intermolecular binding energies. Parameters for the XDM(BJ), XDM(Z), XCDM(BJ), and XCDM(Z) dispersion corrections, optimised for all combinations of eleven density functionals and five basis sets, are included in the ESI. 

Herein, we will focus on only nine representative functionals, using data near the basis-set limit to avoid any conflating effects of basis-set incompleteness errors. At the GGA level of theory, we considered the PBE\cite{becke1986density, perdew1996generalized, perdew1997erratum} and B86bPBE\cite{becke1986large} functionals. At the global hybrid level, we selected six GGA-based hybrids including B3LYP,\cite{becke1988density, beck1993density, lee1988development, stephens1994ab, vosko1980accurate} popular for molecular thermochemistry, PBE0,\cite{adamo1999toward} popular in solid-state chemistry, and our recommended B86bPBE0;\cite{price2023xdm}  we also used their analogues with 50\% exact exchange---BHLYP\cite{becke1993new}, PBE50, and B86bPBE50---which should exhibit reduced delocalisation error.\cite{bryenton2023delocalization} Finally, we also considered the range-separated GGA-based hybrid \mbox{LC-$\omega$PBE} ($\omega = 0.4~\text{bohr}^{-1}$),\cite{vydrov2006assessment, vydrov2007tests} which has previously demonstrated excellent performance on GMTKN55 as a minimally empirical functional when paired with the D3(BJ) dispersion correction.\cite{goerigk2017look} Our focus is limited to methods with simple functional forms as this is consistent with an ``Occam's Razor'' approach to DFA development. We explicitly do not consider any exchange-correlation functionals involving empirically fit parameters, with the exception of B3LYP, which involves three parameters and was fit to the G1\cite{g1a,g1b} thermochemistry set only. We similarly do not consider any double-hybrid functionals due to their reliance on virtual orbitals.

For GMTKN55, all FHI-aims calculations used the \texttt{tight} basis, except for subsets containing anions. HB21, BH76, BH76RC, and G21EA used \texttt{tier2\_aug2} for all atoms; IL16 used \texttt{tier2\_aug2} for all O, F, S, and Cl atoms; and WATER27 used \texttt{tier2\_aug2} for O atoms only for reactions involving anions, as this basis caused linear dependencies in the SCF for some of the larger, neutral water clusters. In all cases, the damping parameters were kept at the same values optimised for the \texttt{tight} basis settings as these are already sufficiently converged as to approach the basis-set limit. As some SCF convergence problems were encountered for the range-separated hybrid functionals in FHI-aims (see the ESI), the \mbox{LC-$\omega$PBE} data were obtained in combination with the \texttt{aug-cc-pVTZ} basis set using the Gaussian16 program,\cite{frisch2016gaussian} with the dispersion corrections applied \textit{ad hoc} using the postg code.\cite{otero2025postgxcdm}

Turning to the solid-state, only FHI-aims calculations were performed and only the two GGA and four global-hybrid functionals were considered (B3LYP and BHLYP were omitted as the asymptotic constraint used in the construction of the B88 exchange functional\cite{becke1988density} is not relevant for solid-state systems). For the X23, ICE13, and HalCrys4 benchmarks, GGA calculations used both \texttt{tight} and \texttt{lightdenser}  basis settings. The latter is our recommended basis for most solid-state calculations, particularly geometry optimisations, although there will be some residual basis-set incompleteness error. As hybrid calculations with the \texttt{tight} basis require prohibitive amounts of memory, only \texttt{lightdenser} calculations were performed. Hybrid results with the \texttt{tight} basis were approximated using an additive basis set correction evaluated at the converged GGA/\texttt{lightdenser} geometries:\cite{hoja2018first,price2023xdm}
\begin{align}\label{eq:bsc}
E(\text{hybrid/\texttt{tight}}) & \approx E(\text{hybrid/\texttt{lightdenser}}) \notag\\
& + E(\text{GGA/\texttt{tight}}) \notag\\
&- E(\text{GGA/\texttt{lightdenser}}) \,.
\end{align}
For the LM26 benchmark (and its LM11 subset), only GGA calculations using the \texttt{lightdenser} 
and \texttt{tight} basis settings were performed. Hybrid results are not reported due to SCF convergence issues, likely arising due to the small band gaps in these semiconducting materials.

Lastly, we highlight the computational efficiency of the XDM-based post-SCF dispersion corrections. As shown in Table~\ref{tab:timings}, these corrections account for only a small fraction of the total CPU time compared to even a single SCF step. The increased overhead to compute XCDM is negligible relative to XDM, and Z-damping is slightly quicker than BJ-damping, although not enough to be significant during a geometry optimisation.

\begin{table}
\caption{
Timing comparisons for XDM and its variants are reported as the mean of the percent van der Waals (vdW) time per system. For the ``1 SCF Step'' column, each GMTKN55 system was reinitialized from a converged SCF using the \texttt{elsi\_restart} feature, and was allowed to converge---typically one SCF step---using the B86bPBE0 hybrid functional and \texttt{tight} basis settings. For the ``1 Opt Step'' column, each system in the X23 benchmark was started from a pre-converged geometry, thus a single geometry optimisation step was calculated. The results combine data from PBE, B86bPBE, and their associated 25\% and 50\% hybrid functionals, all using the \texttt{lightdenser} basis.
}
\centering
\begin{tabular}{l|cc} \hline
Method   & 1 SCF Step & 1 Opt Step \\ \hline
XDM(BJ)  & 9.42\%     &  3.20\%    \\
XDM(Z)   & 9.16\%     &  3.20\%    \\
XCDM(BJ) & 9.44\%     &  3.24\%    \\
XCDM(Z)  & 9.17\%     &  3.28\%    \\ \hline
\end{tabular}
\label{tab:timings}
\end{table}

\section{Results and Discussion}

\subsection{Molecular Benchmarks}

The focus of this section is the GMTKN55 set, comprised of 55 diverse molecular benchmarks. The summarized results are presented below, while full statistics for each benchmark with all functionals and dispersion corrections, as well as the WTMAD values for each category, are provided in the ESI.
A recent study by Becke demonstrated the improved performance of Z-damping compared to BJ-damping for alkali metal clusters in the ALK8 benchmark when paired with a double-hybrid functional.\cite{becke2024remarkably} However, in that work, the Z-damping parameter was fitted to the GMTKN55 itself, which may have introduced a confounding variable. Here, we fitted XDM(Z) to the canonical KB49 set and extended the comparison to a range of common, minimally empirical, density functionals.

\begin{figure}[t!]
\includegraphics[width=\columnwidth]{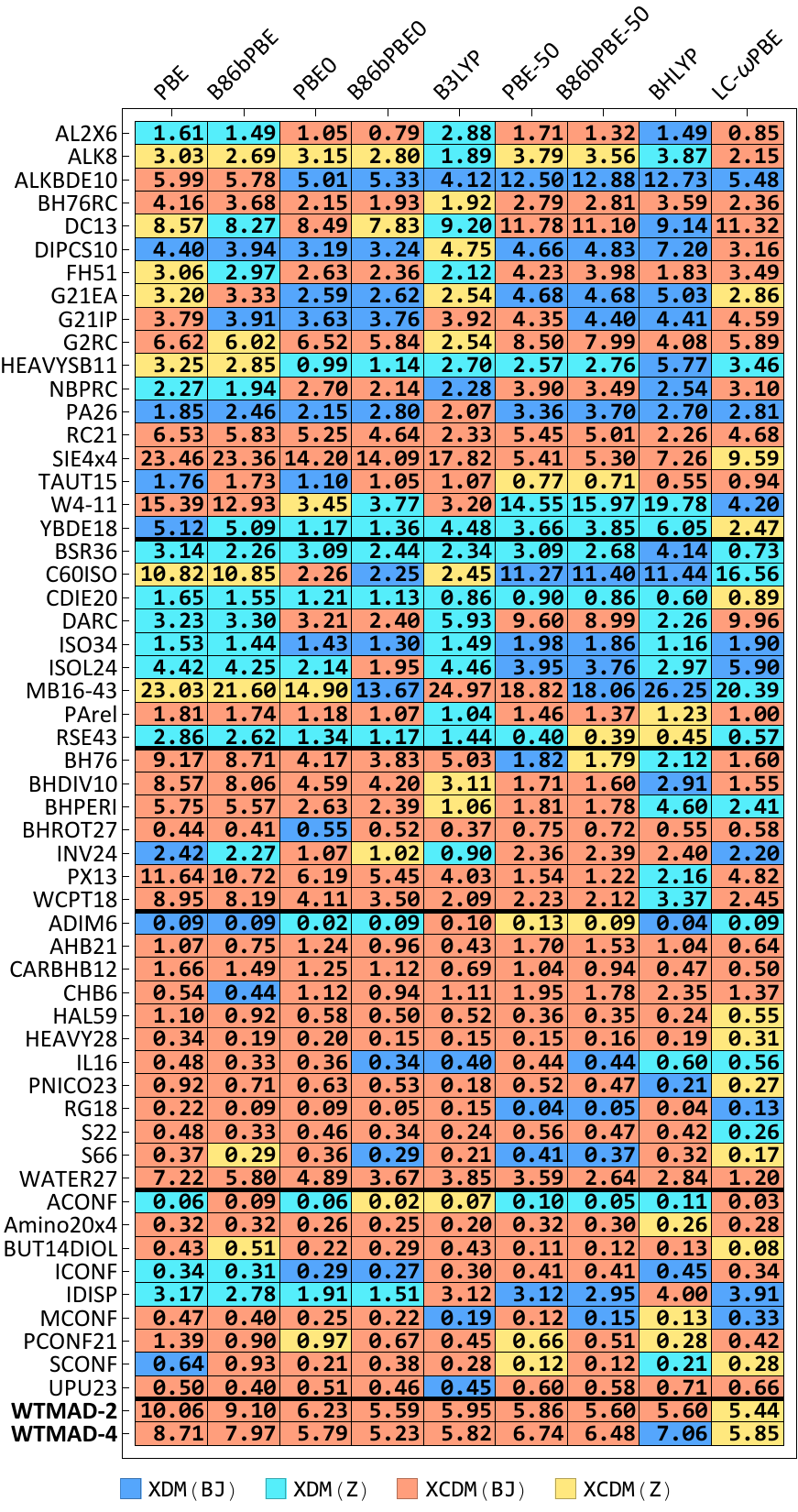}
\caption{Minimum MAEs, in kcal/mol, obtained for each benchmark within the GMTKN55 set for each DFA considered. The dispersion correction yielding this MAE is indicated by the colour. \mbox{WTMAD-2} and \mbox{WTMAD-4} results for the entire benchmark are also shown. Thick black lines partition GMTKN55 into its composite categories: (from top to bottom) ``Basic + Small'', ``Iso + Large'', ``Barriers'', ``Intermolecular NCI'', and ``Intramolecular NCI''.}
\label{fig:gmtkn55combos}
\end{figure}

Figure~\ref{fig:gmtkn55combos} shows the best-performing dispersion correction among XDM(BJ), XDM(Z), XCDM(BJ), and XCDM(Z) for each functional and benchmark alongside the corresponding MAE. While XCDM(BJ) clearly performs better for most systems, Z-damping shows clear improvements for specific cases, including ALK8 (dissociation and other reactions of alkaline compounds), HEAVYSB11 (dissociation energies in heavy-element compounds), YBDE18 (bond-dissociation energies in ylides), and much of the ``Iso+Large'' category (reaction energies for large systems and isomerisation reactions). \mbox{LC-$\omega$PBE} also tends to pair better with Z-damping, although further testing is needed to see if this is broadly applicable to range-separated hybrid functionals.

\begin{figure}[t!]
\includegraphics[width=\columnwidth]{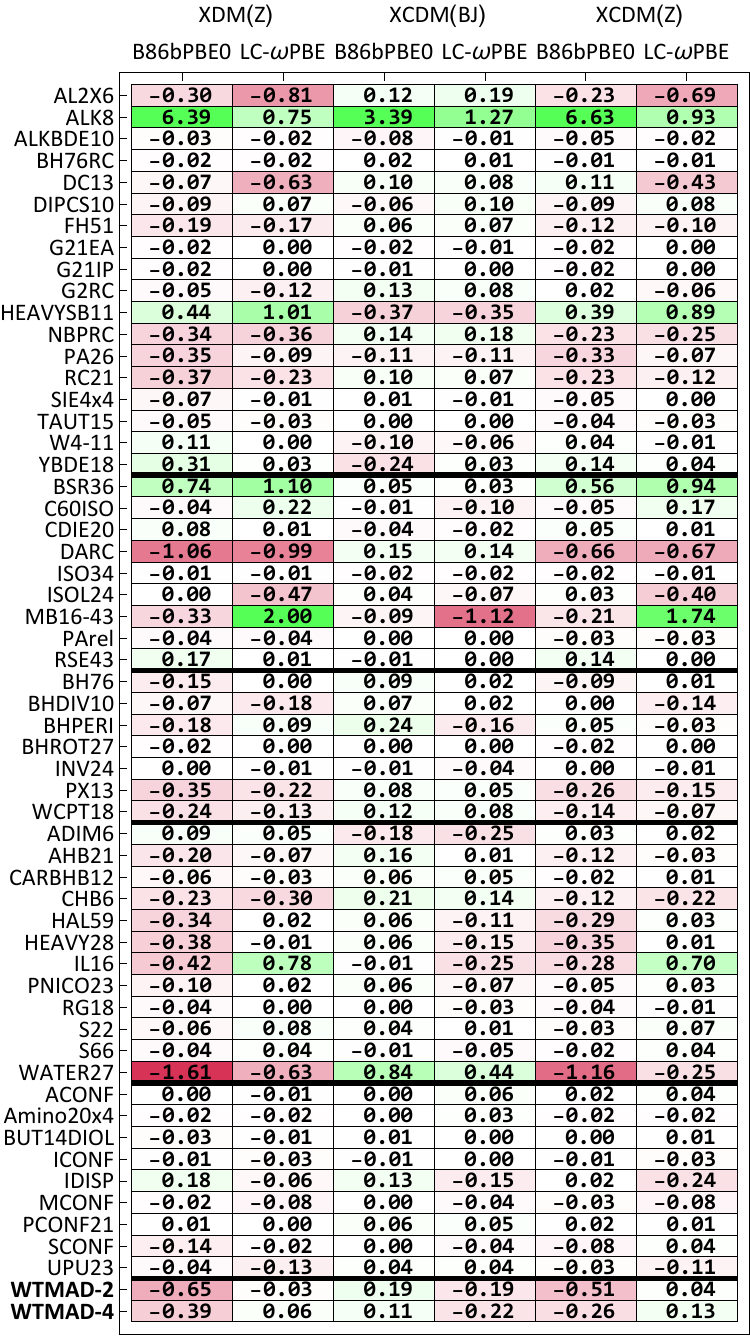}
\caption{MAE differences, in kcal/mol, relative to $\text{XDM(BJ)}$ results for each GMTKN55 benchmark with the three new XDM variants. Data is shown only for the two generally best-performing functionals: B86bPBE0 and LC-$\omega$PBE. The red/green colours indicate increases/decreases in the MAE for a particular data set. The \mbox{WTMAD-2} and \mbox{WTMAD-4} differences for the entire benchmark are also shown. Thick black lines partition GMTKN55 into its composite categories: (from top to bottom) ``Basic + Small'', ``Iso + Large'', ``Barriers'', ``Intermolecular NCI'', and ``Intramolecular NCI''.}
\label{fig:gmtkn55heatmap}
\end{figure}

To quantify the differences between the four dispersion corrections, the MAE difference of XDM(Z), XCDM(BJ), and XCDM(Z) compared to XDM(BJ) for each benchmark is shown in Figure~\ref{fig:gmtkn55heatmap}. The B86bPBE0 and \mbox{LC-$\omega$PBE} functionals were selected for this assessment because B86bPBE0 is representative of most other global hybrids tested, while \mbox{LC-$\omega$PBE} should reduce delocalisation error due to its range separation. As expected, Z-damping alleviates the issues with alkali metal clusters and substantially improves performance on the ALK8 set, lowering the MAE by $>$6~kcal/mol with B86bPBE0. However, the ALK8 set is not an outlier for BJ-damping when paired with LC-$\omega$PBE, indicating that there is considerable interplay between the base functional and dispersion damping for this data set. Additionally, XCDM(BJ) at least partially alleviates the overbinding of ALK8, although not to the same extent as Z-damping. It is also notable that the large improvement in the ALK8 MAE with Z-damping does not carry over to the WTMAD values, likely due to it providing slightly worse performance for most of the other benchmarks. It would perhaps be desirable to consider a weighted root-mean-square deviation, or some other metric that would reward consistency across the subsets while punishing extreme outliers more harshly in its evaluation. 

The largest worsening of performance seen with Z-damping is for the WATER27 set, where the MAD rose by $>$1 kcal/mol with B86bPBE0, showing increased overbinding relative to BJ-damping. This is partially due to delocalisation error in the base functional as WATER27 exhibits cooperative hydrogen bonding, in many cases involving ions; the observed overbinding is consistent with the ICE13 benchmark results in Section~\ref{ss:solids} as well. However, worsening performance with Z-damping, relative to BJ-damping, is also seen for WATER27 with LC-$\omega$PBE. Perhaps the $(Z_i+Z_j)$ term appearing in the Z-damping function (Eq.~\ref{eq:zdamp}) is somewhat too weak for hydrogen, and this could be investigated in future work.

In addition to our four XDM variants, we also evaluated the performance of the MBD family of DCs available in \mbox{FHI-aims} (TS, MBD@rsSCS, and \mbox{MBD-NL}), and compare with literature D3(BJ) results.\cite{goerigk2017look} Due to the limited availability of damping parameters, only the PBE and PBE0 DFAs were considered. WTMAD values and per-category breakdowns are provided in the ESI. Qualitatively, the TS method appears to struggle with the Iso+Large category and D3(BJ) with the reaction barriers, while MBD@rsSCS, MBD-NL, XDM, and XCDM are more consistently accurate across all categories. XCDM(BJ) yields the best results of any of the DCs for the GMTKN55, according to both \mbox{WTMAD-2} and \mbox{WTMAD-4} metrics. 

\begin{table}[t]
\caption{\mbox{WTMAD-2} and \mbox{WTMAD-4} results in kcal/mol for the GMTKN55 benchmark for selected functionals and dispersion corrections.}
\label{tab:gmtkn55}
\centering
\renewcommand{\tabcolsep}{3pt}
\begin{tabular}{l|cccc}\hline
WTMAD-2        & XDM(BJ) & XDM(Z) & XCDM(BJ) & XCDM(Z) \\ \hline
PBE            & 10.40 & 10.78 & 10.06 & 10.70 \\
B86bPBE        & ~9.38 & 10.00 & ~9.10 & ~9.87 \\ \hline
PBE0           & ~6.55 & ~6.96 & ~6.23 & ~6.88 \\
B86bPBE0       & ~5.78 & ~6.43 & \textbf{~5.59} & ~6.29 \\
B3LYP          & ~6.27 & ~6.88 & ~5.95 & ~6.76 \\ \hline
PBE50          & ~6.06 & ~6.44 & ~5.86 & ~6.37 \\
B86bPBE50      & ~5.67 & ~6.20 & ~5.60 & ~6.11 \\ 
BHLYP          & ~5.72 & ~6.09 & ~5.60 & ~6.08 \\ \hline
LC-$\omega$PBE & \textbf{~5.48} & \textbf{~5.51} & ~5.67 & \textbf{~5.44} \\ \hline 
\end{tabular}
\bigskip\\
\begin{tabular}{l|cccc}\hline
WTMAD-4        & XDM(BJ) & XDM(Z) & XCDM(BJ) & XCDM(Z) \\ \hline
PBE            & 8.99    & 9.10   & 8.71     & 9.05    \\
B86bPBE        & 8.13    & 8.52   & 7.97     & 8.42    \\ \hline
PBE0           & 6.06    & 6.12   & 5.79     & 6.06    \\
B86bPBE0       & \textbf{5.34} & \textbf{5.73} & \textbf{5.23} & \textbf{5.60} \\
B3LYP          & 6.12    & 6.44   & 5.82     & 6.37    \\ \hline
PBE50          & 6.92    & 6.95   & 6.74     & 6.90    \\
B86bPBE50      & 6.50    & 6.74   & 6.48     & 6.66    \\
BHLYP          & 7.06    & 7.16   & 7.08     & 7.25    \\ \hline
LC-$\omega$PBE & 5.98    & 5.92   & 6.20     & 5.85    \\ \hline
\end{tabular}
\end{table}

To this point, we have focused on comparing only the various dispersion corrections, but the overall performance for the GMTKN55 is heavily reliant on the choice the underlying base density functional. The \mbox{WTMAD-2} and \mbox{WTMAD-4} results obtained for all nine functionals with the XDM(BJ), XDM(Z), XCDM(BJ), and XCDM(Z) dispersion corrections are collected in Table~\ref{tab:gmtkn55}. As expected, the GGA functionals show larger errors than the hybrid and range-separated hybrid functionals. B86b exchange  generally outperforms PBE exchange, which reinforces our previous conclusion as to the importance of using a dispersionless DFA in combination with post-SCF dispersion corrections.\cite{price2021requirements} The best performing methods overall are \mbox{B86bPBE0-XCDM(BJ)} and {LC-$\omega$PBE-XCDM(Z)}, with the WTMAD-4 favouring the former and WTMAD-2 favouring the latter, due to its greater weighting of the BH76 set. It is particularly notable that B86bPBE0 consistently achieves the minimum error on the MB16-43  ``mindless benchmarking'' set, yielding MAEs of 13.6--14.0 kcal/mol for all four dispersion corrections considered. For comparison, it has been noted that ``MADs for MB16-43 usually exceed 15 kcal/mol for most dispersion-corrected hybrid DFAs.''\cite{goerigk2017look} This strongly indicates that the B86bPBE0 functional, in combination with any XDM or XCDM dispersion correction, captures the relevant physics well. 

While we recommend the new WTMAD-4 going forward, use of the WTMAD-2 metric allows comparison of the results in Table~\ref{tab:gmtkn55} with previous literature.\cite{goerigk2017look,santra2019minimally} Our results on GMTKN55 show consistently strong performance for both GGA-based global hybrids and range-separated hybrids. While lower WTMAD-2 values can be obtained by functionals with 10 or more fit parameters,\cite{goerigk2017look} these have much more complicated functional forms, relying on power-series expansions and either range-separation or meta-GGA ingredients. While fitting no parameters in the base DFAs whatsoever, our results rank 2nd through 6th among all GGA-based global hybrids, surpassed only by \mbox{revPBE0-D3(BJ)/\texttt{def2-QZVPP(D)}}, which achieved a \mbox{WTMAD-2} of 5.43.\cite{santra2019minimally} Despite being slightly higher, the \mbox{WTMAD-2} of 5.59 with \mbox{B86bPBE0-XCDM(BJ)/\texttt{tight}} is notable for a number of reasons. First, the strong performance of D3(BJ) and D4 on GMTKN55 can be partially attributed to the large overlap between their damping parameterisation set (S22, S22+, ACONF, SCONF, PCONF, CCONF, ADIM6, RG6) and GMTKN55 itself, whereas XDM and XCDM are parameterized using the external KB49 set. Second, the \texttt{tight} basis set in FHI-aims includes fewer functions than the typical \texttt{def2-QZVPP(D)} basis for this benchmark (thanks to using numerical atom-centered orbitals, rather than Gaussian-type orbitals), yet delivers comparable performance.\cite{abbott2025roadmap} Finally, B86bPBE0-XCDM(BJ)/\texttt{tight} yields results comparable to the best minimally empirical range-separated hybrids: LC-$\omega$hPBE-D3(BJ)/\texttt{(aug-)def2-QZVP} with a \mbox{WTMAD-2} of 5.56\cite{goerigk2017look} and, now, LC-$\omega$PBE-XCDM(Z)/\texttt{aug-cc-pVTZ} with a \mbox{WTMAD-2} of 5.44.

\subsection{Solid-State Benchmarks}\label{ss:solids}

\begin{table*}[t]
\caption{
Mean absolute errors, in kcal/mol, for the X23, HalCrys4, and ICE13 (absolute and relative) lattice-energy benchmarks. All results are shown for \texttt{tight} basis settings at \texttt{lightdenser} geometries; for the hybrid functionals, this involved the basis-set correction of Eq.~\ref{eq:bsc}.
\label{tab:molcrys}
}
\begin{tabular}{l|cccc|cccc|cccc|cccc}\hline
 &         \multicolumn{4}{c|}{X23} & \multicolumn{4}{c|}{HalCrys4} & \multicolumn{4}{c|}{ICE13-Abs} & \multicolumn{4}{c}{ICE13-Rel} \\
 & \multicolumn{2}{c}{XDM}  & \multicolumn{2}{c|}{XCDM}  & \multicolumn{2}{c}{XDM}  & \multicolumn{2}{c|}{XCDM}  
 & \multicolumn{2}{c}{XDM}  & \multicolumn{2}{c|}{XCDM}  & \multicolumn{2}{c}{XDM}  & \multicolumn{2}{c}{XCDM}  \\
Functional &BJ   &  Z   & BJ   & Z    & BJ   & Z    & BJ   & Z     &  BJ   & Z    & BJ   & Z    & BJ   & Z    & BJ   & Z    \\ \hline
PBE        &1.13 & 0.92 & 0.63 & 0.98 & 5.49 & 4.12 & 5.97 & 4.73  &  1.44 & 2.10 & 1.43 & 2.13 & 0.82 & 0.61 & 0.83 & 0.59 \\
B86bPBE    &0.70 & 0.81 & 0.63 & 1.20 & 4.70 & 5.03 & 5.66 & 5.68  &  1.56 & 1.88 & 1.29 & 1.87 & 0.52 & 0.41 & 0.60 & 0.40 \\ \hline
PBE0       &1.00 & 0.66 & 0.53 & 0.74 & 1.61 & 0.57 & 1.99 & 0.59  &  0.43 & 0.50 & 0.43 & 0.52 & 0.48 & 0.29 & 0.49 & 0.29 \\
B86bPBE0   &0.48 & 0.61 & 0.65 & 1.19 & 1.21 & 0.86 & 1.84 & 1.30  &  0.30 & 0.36 & 0.40 & 0.35 & 0.31 & 0.17 & 0.36 & 0.17 \\ \hline
PBE50     &0.87 & 0.75 & 0.60 & 0.79 & 1.78 & 3.78 & 1.19 & 3.38  &  1.30 & 0.69 & 1.33 & 0.66 & 0.21 & 0.24 & 0.22 & 0.25 \\
B86bPBE50 &0.51 & 0.73 & 0.73 & 1.24 & 1.10 & 3.36 & 0.87 & 2.90  &  1.25 & 0.73 & 1.37 & 0.73 & 0.18 & 0.33 & 0.19 & 0.34 \\ \hline
\end{tabular}
\end{table*}

While both XCDM and Z-damping appear consistently reliable across the GMTKN55, it is crucial to also examine their performance in the solid state. Therefore, we examine the following solid-state benchmarks: the molecular crystal structures of X23, HalCrys4, ICE13-Abs, and ICE13-Rel, as well as the layered materials of LM26. For the molecular crystal structures, tabulated results for XDM(BJ), XDM(Z), XCDM(BJ), and XCDM(Z) using the basis-set correction of Eq.~\ref{eq:bsc} are presented in Table~\ref{tab:molcrys}. 

Conventionally, only results from the largest basis set are reported to avoid confounding variables such as error cancellation. However, as shown in the ESI, the various DCs also perform with exceptional accuracy and consistency for the molecular crystal benchmarks with the \texttt{lightdenser} basis setting, rivalling---or even exceeding---the basis-set-corrected results in Table~\ref{tab:molcrys}. In particular, XCDM(BJ) gives MAEs for the X23 set of 0.50-0.65 kcal/mol across all functionals  considered. This performance is worth noting, as these benchmarks are indicative of a method's effectiveness for crystal structure prediction (CSP). In CSP workflows, basis-set corrections are often used only for final energy refinement due to time and computational constraints; in practice, geometry optimisations and preliminary energy ranking typically employ a smaller basis such as \texttt{lightdenser}.

Looking at the basis-set corrected data for X23 specifically, we see that XDM(BJ), XCDM(Z), and XDM(Z) performed similarly, while XCDM(BJ) performed slightly better. The signed mean errors (shown in the ESI) reveal that XDM(BJ) underbound on average, where XCDM(BJ) and XDM(Z) both shifted the mean error closer to zero. Combining both into XCDM(Z) overcorrected and led to slight overbinding. For the ICE13 and HalCrys4 datasets, hybrid functionals tend to outperform GGAs, due to reduction of  delocalisation error.\cite{bryenton2023delocalization} ICE13, which involves co-operative hydrogen-bonding networks, and HalCrys4, which involves halogen bonding, are paradigmatic examples of systems affected by this error.  While XDM(BJ) and XCDM(BJ) performed  equivalently for HalCrys4, we note that Z-damping performed better for 25\% hybrids, and worse for 50\% hybrids, indicating an interplay between  dispersion binding and delocalisation error in the base DFA. Also, there was an improvement for the relative ICE13 lattice energies for  Z-damping when paired with B86bPBE0 and PBE0, which may be beneficial for polymorph ranking.

\begin{figure}[t]
\centering
\includegraphics[width=\columnwidth]{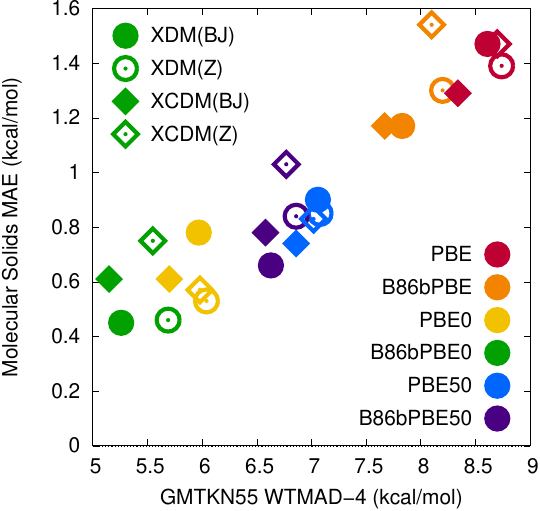}\\
\caption{Scatter plot show the GMTKN55 \mbox{WTMAD-4} values (in kcal/mol) on the $x$-axis, and the overall mean absolute error (MAEs) for the basis-set-corrected solid-state data from Table~\ref{tab:molcrys} on the $y$-axis. XDM is indicated by circles and XCDM by diamonds; solid and hollow shapes denote BJ- and Z-damping respectively. The overall MAE was computed as the mean of all individual reaction errors across all four benchmarks.}
\label{fig:scatter}
\end{figure}

The performance of XDM(Z) with the 25\% hybrid functionals on all four molecular-crystal benchmarks is notable. There is only slight degradation for X23 and ICE13-Abs relative to BJ-damping, a small improvement for ICE13-Rel, and a large net improvement for HalCrys4 (which contains heavier elements). This performance is impressive considering that one empirical parameter was eliminated from the damping function. To unify our GMTKN55 and molecular-crystal results, we have compiled the \mbox{WTMAD-4} and the basis-set-corrected solid-state (X23, ICE13, ICE13-Rel, HalCrys4) data for the PBE and B86bPBE GGA functionals, and their 25\% and 50\% hybrid counterparts Figure~\ref{fig:scatter}.  The results highlight the greater importance of exact-exchange mixing than the choice of dispersion correction. Exactly which XDM variant is the best performing depends on the base functional and benchmark set, but we recommend XDM(Z) as a good general method for both molecular and solid-state applications due to its overall reliability and need for only a single damping parameter.

\begin{table}
\centering
\caption{
Mean absolute errors, in kcal/mol/cell, for the LM26 benchmark and Tawfik's subset, LM11, calculated using the \texttt{tight} basis set in FHI-aims. Analogous results in meV/\AA$^2$ units are given in the ESI.
\label{tab:LM26}
}
\begin{tabular}{l|cccc}\hline
         & \multicolumn{2}{c}{B86bPBE} & \multicolumn{2}{c}{PBE}  \\
Method   & LM11 &  LM26 &  LM11 &  LM26 \\ \hline
XDM(BJ)  & 1.5  &  2.0  &  1.3  &  1.8  \\
XDM(Z)   & 1.4  &  2.0  &  1.0  &  1.4  \\
XCDM(BJ) & 3.6  &  4.2  &  3.3  &  4.0  \\
XCDM(Z)  & 3.3  &  4.0  &  2.5  &  3.1  \\ \hline
\end{tabular}
\end{table}

Finally, we turn to the layered materials benchmark presented in Table~\ref{tab:LM26}, where XDM(Z) yields notable improvements over XDM(BJ), consistent with its good behaviour in the GMTKN55 for metal clusters. Unfortunately the XCDM methods exhibit markedly worse performance. Our analysis reveals that XDM slightly overbinds these layered materials, and this error is exacerbated by the dynamical correlation contributions introduced in XCDM. We attribute this to two main factors. First, XDM relies on a semi-empirical treatment of atom-in-molecule polarizabilities derived via a Hirshfeld partitioning of the electron density. While this approximation is valid for most molecular systems---as evidenced by the excellent performance on MolC6---it tends to overestimate polarizabilities for metals, such as those within the LM26 benchmark. Second, the canonical implementation of the XDM method neglects the Axilrod--Teller--Muto (ATM) three-body dispersion term.\cite{axilrod1943interaction, muto1943force} This omission was intentional due to the ATM term contributing negligibly to the dispersion binding of intermolecular complexes when combined with XDM, coupled with the added computational complexity of summing over atomic trimers.\cite{otero2013many, otero2020many} However, the ATM term is known to be repulsive for equilateral and right-angle atom configurations, while maximally attractive in linear arrangements.\cite{axilrod1943interaction} XCDM captures missing physics by including dynamical correlation, which increases the interatomic attraction. Perhaps the neglect of this correlation previously offset the missing repulsion from the ATM term for these layered materials; this will be the subject of future work.

\section{Summary}

This study improves the dispersion physics of the XDM model, introducing new variants that include dynamical correlation effects and address previous overbinding of metal clusters. It is also the first to test the XDM (and MBD) methods for the GMTKN55 data set, enabling a direct, head-to-head comparison of the most widely-used dispersion corrections on a comprehensive benchmark for general main-group thermochemistry, kinetics, and non-covalent interactions. Additionally, we identified unintended behaviour in previous WTMAD weighting schemes and have introduced \mbox{WTMAD-4} to ensure each benchmark within GMTKN55 is treated fairly. 

All XDM variants proposed and tested here performed extremely well for molecular systems, with the results typically being more sensitive to the choice of base functional than dispersion correction. B86bPBE0 is generally the best exchange-correlation functional among those tested and, despite its simplicity, gives WTMAD values on par with the best minimally empirical global and range-separated hybrids in the literature \cite{goerigk2017look,santra2019minimally} when paired with any of our XDM variants. We attribute the exceptional performance of B86bPBE0 to its adherence to known physical limits.\cite{price2021requirements}

Comparing the dispersion corrections, the canonical XDM(BJ) method showed strong results in all cases with the exception of the ALK8 benchmark, which originally motivated the study into Z-damping. XDM(Z) completely resolved this error and, despite eliminating one empirical parameter, it still performed on par with other leading dispersion corrections on GMTKN55. Notably, it yielded our third lowest \mbox{WTMAD-2} when paired with \mbox{LC-$\omega$PBE}. 
The inclusion of dynamical correlation effects in XCDM eliminated the systematic underestimation of molecular dispersion coefficients, giving improved agreement with available reference data. Consequently, XCDM(BJ) was the most accurate dispersion method tested for molecular systems, providing the lowest WTMAD values when paired with most DFAs considered. However, its  drawback is its poor performance on the layered-material benchmark, LM26, which we attribute to the semi-empirical treatment of XDM polarizabilities causing inflated dispersion coefficients for metals, and possibly to the omission of the Axilrod--Teller--Muto (ATM) three-body term. In contrast, XDM(Z) was consistently accurate for all solid-state benchmarks, including LM26 and LM11. It may be an example of a Pauling point\cite{lowdin1985twenty} for the XDM methods, rarely the best but consistently reliable across the widest range of systems. With only one fit parameter, B86bPBE0-XDM(Z) is an excellent choice for a simple, minimally empirical density functional.

\section*{Acknowledgements}
KRB and ERJ thank the Natural Sciences and Engineering Research Council (NSERC) of Canada for financial support and the Atlantic Computing Excellence Network (ACENET) for computational resources. ERJ additionally thanks the Royal Society for a Wolfson Visiting Fellowship, while KRB thanks the Killam Trust, the Government of Nova Scotia, and the Mary Margaret Werner Graduate Scholarship Fund.
    
\section*{Data Availability Statement}
The data that support the findings of this study are available in the supplementary information.

\section*{Conflicts of Interest}

There are no conflicts to report.

\balance

\bibliographystyle{rsc}
\bibliography{main.bib}

\end{document}